# AGL-1

## The Enterprise AI Governance Layer as a Control Plane for Trusted Enterprise Intelligence

A vendor-neutral reference model for governing retrieval, memory, policy, provenance, observability, and agentic execution in enterprise AI systems

**Roopam W. Sure**
Independent Researcher
June 2026

## Abstract

Enterprise artificial intelligence is moving from isolated experimentation toward operational dependency across copilots, retrieval-augmented generation systems, autonomous agents, and AI-enabled business workflows. As this transition accelerates, the primary enterprise challenge is no longer only model access or inference scale. It is governed intelligence operations: the ability to enforce authorization, preserve contextual lineage, control persistent memory, detect stale or conflicting knowledge, constrain agentic execution, and produce audit-ready evidence across distributed AI estates.

This paper introduces AGL-1, the Enterprise AI Governance Layer, as a vendor-neutral reference model for the control plane that should operate across foundation models, retrieval systems, orchestration frameworks, enterprise memory, policy engines, observability systems, tools, APIs, and business applications. AGL-1 does not assume that governance is absent from modern platforms. On the contrary, recent hyperscaler and enterprise software capabilities show that AI governance is becoming a platform layer. The remaining enterprise problem is how to integrate those capabilities into a coherent operating model across heterogeneous AI environments.

Building on governed knowledge-system principles introduced in GKS-5, AGL-1 generalizes the governance problem from retrieval-specific controls to full AI execution-path governance. It identifies recurring failure modes such as unauthorized retrieval, stale grounding, unmanaged memory, weak provenance, policy drift, fragmented observability, and uncontrolled autonomous execution. It then defines seven governance domains: identity-aware retrieval, policy enforcement, provenance management, memory governance, knowledge integrity monitoring, agentic execution control, and trust observability.

The central claim is that durable enterprise value from AI will increasingly depend on the ability to govern intelligence at scale. In complex enterprises, trust is not a property of the model alone. It is a property of the system around the model: identity, knowledge, policy, memory, tools, human oversight, and evidence working together as a managed control plane.

**Keywords:** enterprise AI governance; AI control plane; agentic AI; retrieval-augmented generation; enterprise memory; policy enforcement; provenance; AI observability; governed intelligence

## Executive Summary

The next phase of enterprise AI will be decided less by who has access to the strongest model and more by who can operate AI systems safely across real business workflows. Many organizations can already build copilots, connect models to documents, and expose AI features inside enterprise applications. Fewer can answer the operational questions that matter when those systems start influencing decisions or taking action: who was allowed to access the context, which policy was applied, what memory was used or written, which tool was called, who approved the action, and what evidence exists after the fact.

AGL-1 frames these concerns as a control-plane problem. The governance layer is not another review board, checklist, or compliance artifact. It is a design-time and runtime layer that coordinates identity, policy, provenance, memory, knowledge integrity, agent permissions, and observability across the AI execution path.

The practical message for CTOs, CIOs, Chief AI Officers, Field CTOs, and enterprise architects is direct: do not treat each AI application as a self-contained governance island. Build a shared governance layer that lets teams move faster because authorization, policy enforcement, evidence production, memory lifecycle, and observability are reusable platform capabilities.

## Table of Contents

# 1. Introduction

Enterprise AI adoption has moved beyond the novelty stage. The first wave was defined by foundation-model access, prompt engineering, copilots, and early retrieval-augmented generation deployments. The next wave is being shaped by a harder question: how should an enterprise govern AI-enabled work across users, models, retrieval systems, persistent memory, autonomous agents, workflow tools, business applications, and regulatory boundaries?

This question changes the architecture discussion. In many organizations, technical teams can already expose a model endpoint, build a chat interface, connect a vector database, automate document summarization, or call an API from an agent. Those capabilities are important, but they do not by themselves answer whether an AI-assisted response or action was authorized, grounded, current, policy-compliant, traceable, auditable, and safe to use.

Enterprise AI programs are therefore moving from an inference-scaling problem to a governance-scaling problem. Inference scalability addresses throughput, latency, model access, cost, and application delivery. Governance scalability addresses access control, source authority, contextual lineage, memory lifecycle, policy propagation, evidence production, model and agent observability, and accountability across distributed AI systems.

This paper proposes that these concerns are converging into a distinct enterprise architecture layer: the Enterprise AI Governance Layer. AGL-1 formalizes that layer as a control plane for trusted enterprise intelligence. It is not a claim that governance tools do not exist. Major cloud and enterprise software providers already expose guardrails, AI risk tooling, agent registries, identity services, policy enforcement, observability, and compliance workflows. The more precise claim is that enterprise leaders need a vendor-neutral reference model for how those capabilities should fit together across a heterogeneous AI estate.

The intended audience includes enterprise AI leaders, CIO and CTO organizations, Field CTOs, Client CTOs, Chief AI Officers, responsible AI leaders, AI platform teams, risk leaders, and enterprise architects responsible for moving generative AI, RAG, and agentic systems from pilots into governed production use.

## 1.1 Contributions

1. Defines the Enterprise AI Governance Layer as a distinct architectural control plane for governing retrieval, memory, policy, provenance, observability, and agentic execution across enterprise AI systems.
2. Proposes AGL-1, a vendor-neutral reference model that decomposes the control plane into operational domains, enforcement points, and evidence-producing governance services.
3. Introduces a governance failure-mode taxonomy that separates model-performance failures from failures of authorization, contextual integrity, memory lifecycle, policy enforcement, auditability, and autonomous execution control.
4. Provides an implementation roadmap and maturity-assessment questions for enterprise leaders who need to convert AI governance from policy aspiration into operating capability.
5. Explains how enterprise technology leadership must adapt ownership structures, platform strategy, and accountability models as AI systems gain memory, tool access, and agency.

# 2. From GKS-5 to AGL-1

GKS-5 argued that many enterprise AI failures are not simply model failures; they are failures in the knowledge system surrounding the model. It focused on governed knowledge systems for retrieval-augmented intelligence, including source authority, metadata discipline, authorization-aware retrieval, freshness management, evaluation, and traceable context assembly.

AGL-1 extends that argument to the broader enterprise AI operating environment. Retrieval remains essential, but production AI systems increasingly include persistent memory, autonomous agents, workflow tools, third-party SaaS integrations, data-loss controls, identity systems, policy engines, and observability pipelines. These systems are no longer isolated knowledge assistants. They are becoming distributed operational infrastructure.

The relationship between the two models can be summarized as follows: GKS-5 defines how enterprise knowledge should be governed for AI retrieval. AGL-1 defines how the broader enterprise AI environment should be governed across retrieval, memory, orchestration, policy enforcement, agentic execution, observability, and enterprise accountability.

*Table 1. Relationship between GKS-5 and AGL-1*

| Model | Primary scope | Core question |
|---|---|---|
| GKS-5 | Governed enterprise knowledge systems and retrieval-augmented intelligence | Can the AI system retrieve and use enterprise knowledge in a controlled, fresh, authorized, and traceable way? |
| AGL-1 | Enterprise AI governance control plane across models, retrieval, memory, agents, tools, applications, and policy | Can the enterprise govern AI-enabled intelligence operations across the full execution path? |

This distinction matters for executive technology leadership. An enterprise may deploy a well-designed RAG pattern and still fail to govern agent actions, memory persistence, delegated authority, cross-application access, policy drift, or audit evidence. AGL-1 treats those concerns as one control-plane problem rather than as disconnected implementation details.

## 3. Industry Signals: Governance Is Becoming a Platform Layer

The AI Governance Layer is not an abstract future concept. The market is already assembling its building blocks. Hyperscalers and enterprise software providers increasingly expose capabilities for AI guardrails, agent identity, agent registries, gateway-mediated policy enforcement, observability, risk management, and enterprise compliance. The pattern is visible even though implementations remain vendor-specific and fragmented.

NIST's AI Risk Management Framework and its Generative AI Profile reinforce the same operating principle from a risk perspective: organizations need repeatable mechanisms to govern, map, measure, and manage AI risks, including risks introduced by generative AI systems [4], [5]. The platform market is now turning parts of that operating principle into technical control surfaces.

AWS provides governance signals through Amazon Bedrock Guardrails and Amazon Bedrock AgentCore. AgentCore includes services such as runtime, memory, gateway, identity, and observability for operating agents and tools in production environments [11]-[14]. Microsoft has explicitly described Agent 365 as a control plane for AI agents and has introduced agent-specific identity and governance concepts through Microsoft Entra Agent ID [15], [16]. Google Cloud's Gemini Enterprise Agent Platform exposes Agent Registry, Agent Gateway, Agent Identity, Model Armor, policy enforcement, observability, RAG-related integration, and governance controls for enterprise agent platforms [18]-[20]. IBM watsonx.governance emphasizes connected governance context through a Governance Graph that maps AI assets, policies, risks, controls, and regulatory requirements across platforms and production environments [21], [22].

These industry signals are important for AGL-1. The paper does not claim that hyperscalers lack AI governance. The claim is that governance capabilities are becoming a platform layer, while enterprise leaders still need a vendor-neutral operating model for how those capabilities fit together across multi-cloud, SaaS, open-source, and internally built AI estates.

*Table 2. Platform signals supporting the AGL-1 thesis*

| Provider signal | Representative capabilities | Implication for AGL-1 |
|---|---|---|
| AWS | Guardrails, AgentCore runtime, memory, gateway, identity, observability, tool integration, and policy controls | Governance is moving into agent and model operations rather than remaining only in review workflows. |
| Microsoft | Agent 365 control plane, Entra Agent ID, lifecycle management, access control, telemetry, compliance, and audit concepts | AI agents require inventory, ownership, identity, limits, monitoring, and stop mechanisms. |
| Google Cloud | Agent Registry, Agent Gateway, Agent Identity, Model Armor, IAM policies, observability, and agent governance | Agentic platforms are converging around gateway-mediated governance and identity-aware execution. |
| IBM | Governance Graph, AI asset inventory, policy/risk/control mapping, regulatory alignment, and cross-platform governance | AI governance increasingly requires connected enterprise context, lineage, controls, and risk relationships. |

# 4. The Enterprise AI Governance Gap

Despite rapid platform progress, many enterprises still operate fragmented AI environments. Departments deploy copilots, vector databases, prompt libraries, model endpoints, SaaS AI features, and agentic workflows independently. This creates an unmanaged AI estate in which the enterprise may not consistently know what systems exist, which data they access, which policies govern them, which users or agents act through them, or what evidence they produce.

The governance gap is not just a compliance gap. It is an operating gap. AI systems synthesize outputs from evolving context, user instructions, model behavior, tool results, historical memory, and policy constraints. When those inputs are distributed and weakly governed, enterprises face new classes of operational risk.

## 4.1 A simple enterprise scenario

Consider a customer-support agent inside a regulated enterprise. The agent receives a complaint, searches internal policy, summarizes the recommended action, opens a case in a workflow system, and sends a response for human review. Nothing in that flow requires science fiction. Each capability exists today. The risk is that the agent may retrieve a superseded policy, use a memory fragment written by a different team, call a tool with broader permissions than the human user, omit an approval step, or leave behind logs that cannot reconstruct what happened.

In that scenario, the model may have behaved exactly as instructed. The failure is not necessarily hallucination. The failure is the absence of a governed execution path: no reliable entitlement check at retrieval time, no policy decision record, no memory version record, no source authority signal, no agent action trace, and no audit bundle. AGL-1 is designed to address precisely this class of failure.

## 4.2 Governance failure modes

*Table 3. Governance failure modes addressed by AGL-1*

| Failure mode | Description | Enterprise consequence |
|---|---|---|
| Unauthorized retrieval | A RAG system retrieves semantically relevant information without enforcing user entitlement, geography, role, contractual access boundaries, or business purpose. | Sensitive data exposure, compliance breach, loss of customer or employee trust. |
| Stale contextual grounding | AI responses are grounded in outdated embeddings, expired policies, superseded procedures, or obsolete knowledge artifacts. | Incorrect advice, operational error, regulatory misalignment. |
| Unmanaged memory | Persistent AI memory accumulates without versioning, expiration, provenance, correction, consent, or rollback controls. | Long-lived error propagation, privacy risk, inconsistent personalization, audit gaps. |
| Weak provenance | The system cannot reconstruct which sources, prompts, policies, memories, tools, or intermediate steps influenced a response or action. | Unverifiable decisions, weak incident response, low executive confidence. |
| Policy drift | Policy constraints are documented but not propagated into retrieval, prompts, orchestration, tools, and agent runtime. | AI behavior diverges from enterprise risk boundaries. |
| Uncontrolled agentic execution | Agents invoke tools, APIs, workflows, or downstream systems without sufficient boundary controls, transaction limits, or approval gates. | Autonomous operational errors, financial exposure, unauthorized actions. |
| Fragmented observability | Model logs, retrieval logs, tool traces, agent trajectories, user feedback, and business outcomes are monitored separately or not monitored at all. | Leaders cannot assess risk, quality, adoption, or incident causality. |

These failure modes are often misclassified as model issues. A hallucinated answer may be caused by model behavior, but it may also arise from stale context, weak retrieval validation, poor source authority, missing policy constraints, or unmanaged memory. A failed agent task may not indicate poor reasoning; it may indicate weak tool governance, missing approval checkpoints, incomplete identity delegation, or insufficient observability.

AGL-1 therefore treats trustworthy enterprise AI as a system property. Trust is not produced by the model alone. It emerges from the interaction of model behavior, governed knowledge, identity, policy, runtime controls, memory lifecycle, human oversight, and evidence records.

# 5. AGL-1 Reference Model

AGL-1 defines the Enterprise AI Governance Layer as a design-time and runtime control plane for trusted enterprise intelligence. The layer mediates interactions among users, agents, retrieval systems, enterprise memory, foundation models, orchestration frameworks, tools, policy systems, observability systems, and business applications.

The purpose of the governance layer is not to slow AI delivery. Its purpose is to make scaled AI delivery possible. Without a shared governance control plane, each AI application must independently solve access control, policy alignment, provenance, memory lifecycle, tool authorization, observability, and audit evidence. That pattern does not scale across large enterprises.

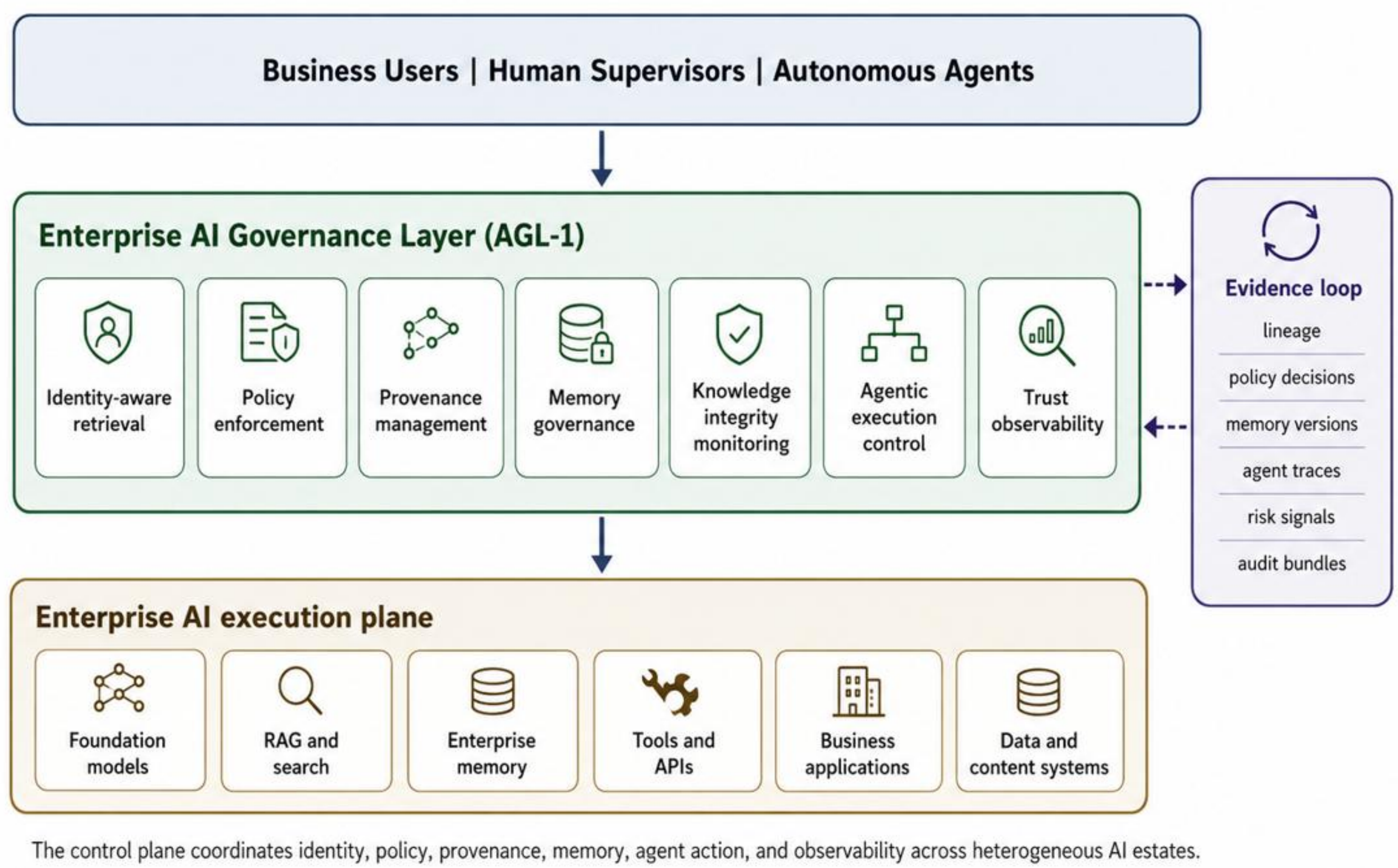

*Figure 1. AGL-1 conceptual control-plane view. The governance layer coordinates identity, policy, provenance, memory, agent action, and observability across heterogeneous enterprise AI execution environments.*

## 5.1 Design principles

1. Governance must be embedded into the AI execution path, not applied only as a post-deployment review.
2. Authorization must be enforced at the retrieval, memory, tool, and agent-action layers, not only at the user-interface layer.
3. Enterprise memory must be attributable, versioned, expirable, correctable, and auditable.
4. AI outputs and agent actions must produce evidence records that can be reconstructed after the fact.
5. Policy must be machine-enforceable at runtime and traceable to enterprise risk intent.
6. Observability must connect technical traces to business, risk, and operational outcomes.

7. The governance model must remain vendor-neutral because enterprise AI estates will span multiple clouds, SaaS platforms, models, data systems, and agent frameworks.

## 5.2 AGL-1 control domains

*Table 4. AGL-1 control domains*

| AGL-1 domain | Enterprise question answered | Control function |
|---|---|---|
| Identity-aware retrieval | Is this user, agent, or workflow allowed to retrieve and use this knowledge? | Enforces entitlement, role, attribute, regional, contractual, and contextual access boundaries. |
| Policy enforcement | Is this AI output, action, or tool invocation allowed under enterprise policy? | Applies business, legal, compliance, safety, and risk constraints before or during execution. |
| Provenance management | Where did this answer, recommendation, or action come from? | Maintains source, version, transformation, prompt, retrieval, model, memory, and tool lineage. |
| Memory governance | Should this memory persist, expire, be corrected, or be rolled back? | Governs lifecycle, retention, consent, correction, versioning, and rollback for persistent AI memory. |
| Knowledge integrity monitoring | Is the system using stale, conflicting, low-authority, or orphaned context? | Detects freshness issues, embedding drift, policy conflicts, source inconsistencies, and index gaps. |
| Agentic execution control | What can this agent do, when does it need approval, and how can it be stopped? | Constrains tools, APIs, delegated authority, escalation, transaction limits, and rollback paths. |
| Trust observability | Can the enterprise monitor, audit, and improve AI behavior across systems? | Produces logs, traces, risk signals, quality metrics, evidence records, and audit bundles across AI workflows. |

AGL-1 is not a single product category. In practice, it may be implemented through combinations of cloud services, data governance platforms, identity providers, policy engines, AI gateways, model platforms, observability tools, compliance systems, and internal enterprise controls. The architectural point is not vendor consolidation. The point is control-plane coherence.

# 6. Core Governance Domains

## 6.1 Identity-aware retrieval

Many enterprise RAG systems begin as relevance engines: retrieve the documents most semantically similar to a query, pass context to a model, and generate an answer. That pattern is insufficient for governed enterprise use. In production environments, retrieval must be constrained by identity, role, data classification, geography, customer relationship, contractual permissions, time-bound entitlements, and contextual business purpose.

Identity-aware retrieval converts retrieval from a relevance-only operation into a policy-constrained operation. The system must evaluate not only whether a document is relevant, but whether the current user, agent, or workflow is allowed to retrieve and use that document for the requested purpose. This control is especially important where vector similarity may surface sensitive information that would not be visible through traditional application permissions.

Minimum evidence: retrieval query, user or agent identity, entitlement decision, source IDs, source versions, and reason code for allow, deny, or redact.

## 6.2 Policy enforcement

Enterprise AI policies often begin as human-readable documents. AGL-1 requires progression from documented policy to enforceable policy. Governance policies should be represented in forms that can constrain prompts, retrieval, context assembly, model routing, tool use, agent plans, approval gates, and output generation.

Policy enforcement should operate before, during, and after AI execution. Before execution, it can determine whether a use case, user, data source, model, or tool is allowed. During execution, it can block unsafe context, restrict tool calls, require approval, or modify routing. After execution, it can generate evidence records, trigger alerts, and support audit review.

Minimum evidence: policy identifier, policy version, evaluated attributes, decision result, decision timestamp, and exception or approval record where applicable.

## 6.3 Provenance management

Provenance management answers a fundamental enterprise question: how did the system arrive at this output or action? For AI systems, provenance extends beyond document citation. It includes source identity, source version, retrieval query, ranking behavior, transformations, prompts, model identity, policy decisions, memory reads, memory writes, tool calls, human approvals, and final outputs.

Without provenance, enterprises cannot reliably investigate AI incidents, defend AI-assisted decisions, improve system quality, or satisfy audit expectations. Provenance should therefore be treated as a first-class architecture concern, not an optional logging feature.

Minimum evidence: source lineage, model and prompt metadata, policy and retrieval decisions, memory events, tool trace, human approval state, and final output hash or record locator.

## 6.4 Enterprise memory governance

Persistent memory changes the enterprise AI risk profile. A stateless chatbot may produce a wrong answer once. A system with unmanaged memory can preserve, reuse, and amplify wrong or sensitive information across sessions, users, teams, or agents. Memory can improve productivity and personalization, but only if it is governed.

Memory governance requires attribution, versioning, retention rules, expiration, correction, access control, review workflows, consent where applicable, and rollback. Enterprises should distinguish between session memory, user memory, team memory, enterprise knowledge memory, and agent memory. Each memory type has different governance requirements.

Minimum evidence: memory owner, source, type, write authority, access scope, retention rule, correction history, expiration date, and rollback reference.

## 6.5 Knowledge integrity monitoring

Enterprise knowledge changes continuously. Policies are revised, products evolve, regulations change, procedures are replaced, and business rules are updated. AI systems that rely on stale embeddings, outdated retrieval indexes, or obsolete knowledge artifacts can produce confident but operationally incorrect responses.

Knowledge integrity monitoring addresses this risk by tracking freshness, source authority, conflicting artifacts, superseded documents, embedding age, index coverage, and orphaned memory. In governed AI systems, knowledge freshness becomes an operational metric rather than a manual content-management concern.

Minimum evidence: source authority score, freshness timestamp, index coverage signal, conflict indicator, supersession relationship, and drift or staleness alert.

## 6.6 Agentic execution control

Agentic AI expands governance requirements because agents do not only generate language; they can plan, call tools, invoke APIs, coordinate workflows, and trigger operational actions. The governance problem therefore shifts from output review to action control.

Agentic execution control should include tool allowlists and denylists, delegated authority rules, approval checkpoints, budget or transaction limits, context constraints, action provenance, interrupt mechanisms, escalation paths, rollback procedures, and post-action audit trails. As agents operate across applications, the enterprise must govern not only what they know, but what they can do.

Minimum evidence: agent identity, plan steps, tool calls, input/output parameters, approval state, execution result, exception handling, and rollback or remediation outcome.

## 6.7 Trust observability

Traditional observability focuses on infrastructure and application health. AI observability must also monitor retrieval quality, grounding, policy decisions, model behavior, agent trajectories, memory operations, user feedback, risk signals, and business outcomes. AGL-1 treats trust observability as a leadership capability because executives cannot scale what they cannot see.

The goal is not merely to collect logs. The goal is to produce actionable governance intelligence: which AI systems exist, who owns them, what data they use, what policies constrain them, how they behave, what incidents occur, what evidence is available, and where risk is accumulating.

Minimum evidence: inventory coverage, control effectiveness, policy violation rate, retrieval quality, grounding signal, drift indicator, incident trend, adoption metric, and business outcome linkage.

## 7. Governance Artifacts and Evidence Records

A governance layer must produce durable evidence. Without evidence records, governance remains aspirational. AGL-1 defines a set of artifacts that should be produced across AI workflows and retained according to the risk tier, regulatory context, and business criticality of the system.

*Table 5. AGL-1 governance artifacts and evidence records*

| Artifact | Purpose | Example contents |
|---|---|---|
| AI system inventory record | Establishes what AI systems, agents, models, and workflows exist and who owns them. | System owner, business purpose, model provider, data sources, risk tier, approved uses, deployment status. |
| Retrieval lineage record | Shows which knowledge artifacts were retrieved and why. | Query, user or agent identity, source IDs, source versions, access decision, ranking metadata, redaction state. |
| Policy decision record | Captures how governance rules were evaluated. | Policy version, evaluated attributes, allow/block/modify decision, reason code, timestamp, exception owner. |
| Memory version record | Tracks persistent memory lifecycle. | Memory owner, source, version, retention, correction history, expiration rule, access scope, rollback pointer. |
| Agent action trace | Reconstructs agent planning and execution. | Plan steps, tool calls, approvals, API actions, intermediate outputs, human interventions, completion state. |
| Risk and quality signal | Supports monitoring and continuous improvement. | Grounding score, policy violation signal, drift indicator, user feedback, exception rate, incident linkage. |
| Audit bundle | Provides a consolidated record for review, incident response, or regulatory inquiry. | Relevant lineage, policies, logs, traces, approvals, outputs, remediation notes, business owner attestation. |

These artifacts serve different stakeholders. Engineers use them to debug systems. Risk teams use them to assess controls. Business owners use them to understand adoption and accountability. Executives use them to decide whether AI systems can be scaled, constrained, retired, or expanded.

The evidence model is also what separates a governance layer from a governance narrative. A policy that cannot be traced to an execution decision is weak. A control that cannot produce evidence is difficult to defend. A system that cannot explain its retrieval, memory, policy, and action path should not be treated as production-grade in high-risk workflows.

## 8. Implementation Roadmap: What a CTO Should Do Next

A reference model is only useful if it changes operating behavior. The following roadmap translates AGL-1 into practical sequencing for enterprise technology leaders. It assumes the organization already has active AI pilots or early production systems and now needs to move toward governed scale.

*Table 6. AGL-1 implementation roadmap*

| Phase | Primary CTO action | What good looks like |
|---|---|---|
| 0-30 days: Inventory and risk tiering | Create an AI system and agent inventory. Identify business owners, data sources, models, tools, integrations, and risk tiers. | Leadership can see what AI systems exist, who owns them, what they touch, and which require immediate governance attention. |
| 31-60 days: Control the highest-risk paths | Prioritize identity-aware retrieval, tool authorization, and policy enforcement for the most sensitive workflows. | High-risk systems enforce access decisions at retrieval and action time, not only at the UI layer. |
| 61-90 days: Produce evidence | Implement retrieval lineage, policy decision records, memory version records, and agent action traces for priority systems. | Risk, engineering, and business owners can reconstruct AI-assisted decisions and actions after the fact. |
| 90-180 days: Govern memory and agents | Define memory classes, retention rules, correction flows, tool limits, approval gates, escalation paths, and kill-switch procedures. | Agents and memory-bearing systems have clear boundaries, operating authority, and rollback procedures. |
| 180+ days: Build trust observability | Create dashboards connecting adoption, control effectiveness, policy exceptions, drift, incidents, quality, and business outcomes. | Executives can decide which AI systems to scale, constrain, remediate, or retire based on evidence rather than anecdotes. |

### 8.1 Minimum viable AGL-1

The minimum viable governance layer should not attempt to solve every governance problem at once. It should establish reusable controls for the riskiest execution paths first. A practical minimum viable AGL-1 includes:

- an AI inventory with named business and technical owners;
- risk tiers for AI systems, agents, data sources, and tool actions;
- identity-aware retrieval for sensitive enterprise knowledge;
- policy enforcement for prohibited content, restricted data, and high-risk actions;
- retrieval lineage and policy decision records;
- agent tool allowlists, approval gates, and interrupt procedures;
- basic trust observability across adoption, exceptions, drift, and incidents.

This is the Monday-morning starting point. The organization does not need a perfect governance platform before scaling AI. It does need a common control model that prevents every team from reinventing partial governance in isolation.

## 9. AGL-1 Maturity Assessment Questions

The following questions help leaders assess whether governance exists as policy language only, as isolated project controls, or as an enterprise control plane. They can be used as a lightweight maturity assessment before deeper mapping to frameworks such as NIST AI RMF, ISO/IEC 42001, sector-specific regulation, and internal technology risk standards.

*Table 7. AGL-1 maturity assessment questions*

| Domain | Maturity question | Weak signal | Strong signal |
|---|---|---|---|
| Inventory | Do we know which AI systems, agents, models, tools, and workflows exist? | Inventory depends on surveys or ad hoc reporting. | Inventory is maintained as part of platform onboarding and release governance. |
| Identity | Can retrieval, memory, and tool access be constrained by user, agent, role, geography, and purpose? | Permissions are enforced mainly at the application UI. | Entitlements are evaluated inside retrieval, memory, tool, and agent execution paths. |
| Policy | Are policies machine-enforceable at runtime? | Policies exist as documents and training material. | Policies produce allow, deny, modify, escalate, or approval decisions during execution. |
| Provenance | Can we reconstruct how a response or action was produced? | Teams rely on partial logs and screenshots. | Source, prompt, retrieval, model, memory, policy, tool, and approval records are linked. |
| Memory | Can persistent AI memory be corrected, expired, scoped, and rolled back? | Memory is treated as an implementation detail. | Memory types have owners, retention rules, access scopes, correction flows, and version records. |
| Agent control | Can agents be limited, interrupted, escalated, and audited? | Agents inherit broad tool permissions or informal approval patterns. | Agents have specific identities, tool limits, approval gates, transaction boundaries, and action traces. |
| Observability | Can executives see risk, quality, adoption, incidents, and control effectiveness? | Telemetry is fragmented across model, app, and infrastructure tools. | Trust observability connects AI traces to business, risk, and operational outcomes. |

A useful scoring approach is simple: 0 means not visible; 1 means documented but manual; 2 means implemented in selected systems; 3 means implemented as a reusable enterprise control. The goal is not to score high everywhere immediately. The goal is to identify which governance gaps create the greatest enterprise exposure.

## 10. Enterprise Leadership and Operating Model Implications

The emergence of an AI Governance Layer has significant implications for enterprise technology leadership. As generative AI systems become embedded in knowledge work, customer operations, software delivery, compliance workflows, and autonomous business processes, governance can no longer remain isolated within policy, risk, or compliance functions. It must become part of the enterprise AI operating model.

CIOs, CTOs, Chief AI Officers, Field CTOs, Client CTOs, enterprise architects, AI platform leaders, and responsible AI leaders will increasingly need to manage AI systems as distributed operational infrastructure. This requires clear ownership of model access, retrieval governance, enterprise memory, policy enforcement, observability, auditability, and agentic execution controls. Without this ownership model, organizations may deploy AI capabilities faster than they can govern them.

The AI Governance Layer therefore represents more than a technical architecture. It is an enterprise control model. It defines how organizations coordinate business intent, technology execution, risk boundaries, and operational accountability across AI-enabled systems. In this model, governance is not a deployment gate at the end of an AI project. It is a runtime capability embedded into how enterprise intelligence is accessed, composed, executed, monitored, and improved.

*Table 8. Operating model ownership for AGL-1*

| Capability | Likely enterprise owners | Leadership concern |
|---|---|---|
| AI system inventory | AI platform team, enterprise architecture, risk office | Do we know what AI systems and agents exist? |
| Identity-aware retrieval | Identity team, data governance, AI platform team | Can AI access only what the user or agent is entitled to use? |
| Policy enforcement | Risk, legal, compliance, platform engineering | Are business and regulatory constraints enforced at runtime? |
| Enterprise memory governance | Data governance, AI platform, privacy, business owners | Can persistent AI memory be trusted, corrected, expired, and audited? |
| Agentic execution control | Platform engineering, security, business application owners | Can agents act safely across tools, APIs, and workflows? |
| Trust observability | SRE, AI platform, risk analytics, business owners | Can leaders see quality, risk, adoption, incidents, and control effectiveness? |

This shift expands the role of enterprise technology leadership. Future AI leaders will need to combine platform strategy, risk architecture, business transformation, data governance, and operational trust engineering. The organizations that succeed will be those that treat governed AI not as a compliance obligation, but as a foundation for scalable enterprise intelligence.

# 11. Strategic Implications for Regulated Enterprises

AGL-1 is especially relevant for regulated and knowledge-intensive industries, including financial services, healthcare, telecommunications, energy, government, insurance, and manufacturing. These industries depend on trusted knowledge, controlled access, regulatory alignment, and operational accountability. They are also likely to adopt AI agents for customer operations, employee productivity, compliance workflows, IT operations, sales enablement, and software delivery.

## 11.1 Financial services

A bank may use AI assistants for policy interpretation, customer-service support, wealth-advisor enablement, fraud operations, and internal software engineering. In this environment, authorization-aware retrieval, source authority, policy enforcement, and audit evidence are mandatory. An answer grounded in outdated policy or unauthorized customer information can create compliance and reputational exposure.

## 11.2 Healthcare

Healthcare organizations face sensitive data, complex access boundaries, evolving clinical and administrative policies, and high consequences for incorrect information. A governance layer can help constrain retrieval, separate clinical and non-clinical use cases, maintain provenance, and enforce approval or escalation points for high-risk workflows.

## 11.3 Telecommunications and customer operations

Telecommunications providers operate large customer-service, network, field-service, billing, and sales-support environments. AI agents may assist employees or customers across these workflows. Governance must manage customer identity, entitlement, tool access, memory, handoffs, escalation, and action traces across complex operational systems.

## 11.4 Government and public sector

Government agencies require transparency, public accountability, records management, security, policy consistency, and traceability. AGL-1 is relevant because public-sector AI systems must not only work; they must be explainable, controllable, evidence-producing, and aligned with policy boundaries.

### 11.5 Enterprise technology strategy

Across industries, the strategic question is shifting from 'which model should we use?' to 'what governed AI operating model should we build?' Model choice remains important, but it is not sufficient. Enterprises will increasingly compete on their ability to deploy AI systems that are trusted enough to operate at scale in real workflows.

## 12. Limitations and Future Work

AGL-1 is a reference model, not a product specification or empirical benchmark. It does not prescribe a single vendor architecture, implementation stack, policy language, observability platform, or risk-scoring method. It also does not claim that all enterprises require the same level of governance maturity for every AI use case.

The model is intentionally high-level because enterprise AI estates are heterogeneous. A large bank, a hospital network, a telecom provider, a manufacturing firm, and a public-sector agency will implement governance differently. However, each will need some way to manage identity-aware retrieval, policy enforcement, provenance, memory lifecycle, agentic execution control, and trust observability.

Future work should convert AGL-1 into more detailed implementation patterns and evaluation methods. Potential extensions include:

- a maturity model for enterprise AI governance layers;
- a scenario-based evaluation framework for governed RAG and agentic AI;
- a reference telemetry schema for AI provenance, retrieval lineage, memory events, and agent traces;
- a control catalog mapped to NIST AI RMF, ISO/IEC 42001, enterprise risk frameworks, and sector-specific regulatory expectations;
- a comparative analysis of hyperscaler, SaaS, and open-source governance-layer implementations;
- reference implementation blueprints for identity-aware retrieval, memory lifecycle governance, and agent action traces.

## 13. Conclusion

Enterprise AI is entering a phase where governance becomes inseparable from platform strategy. The first wave of adoption emphasized model access, prompt engineering, copilots, and experimental RAG. The next wave will require governed intelligence operations across models, retrieval, memory, agents, tools, policies, and business workflows.

AGL-1 defines the Enterprise AI Governance Layer as a control plane for that transition. It formalizes an emerging industry pattern: AI systems are becoming operational infrastructure, and operational infrastructure requires identity, policy, provenance, memory controls, observability, and auditability.

The central claim is not that governance replaces model innovation. Model capability will continue to matter. The claim is that durable enterprise value will increasingly depend on the ability to govern AI systems at scale. In complex enterprises, trusted intelligence is not a property of the model alone. It is a property of the system around the model.

Organizations that build this governance capability will be better positioned to move from pilots to production, from productivity experiments to operational workflows, and from isolated AI applications to governed enterprise intelligence platforms. In that sense, the AI Governance Layer may become one of the defining enterprise architecture layers of the next AI adoption cycle.